\documentclass[10pt]{article}
\usepackage[dvips]{color}
\usepackage{epsfig}
\usepackage{amsmath}
\usepackage{graphicx}

\begin{document}
\setcounter{page}{1}

\pagestyle{plain} \vspace{1cm}
\begin{center}
{\Large \bf Interacting Quintessence in a New Scalar-Torsion Gravity}\\
\small \vspace{1cm} {\bf Behnaz Fazlpour
 \footnote{b.fazlpour@umz.ac.ir}} \\
\vspace{0.5cm} {\it Department of Physics, Babol Branch, Islamic
Azad University, Babol, Iran\\}
\end{center}
\vspace{1.5cm}
\begin{abstract}
In the framework of teleparallel equivalent of general relativity,
we study a gravity theory where a scalar field beyond its minimal
coupling, is also coupled with the vector torsion through a
non-minimal derivative coupling. After a suitable choice of
auxiliary variables that allows us to perform the phase-space
analysis of the model we obtain the critical points and their
stability. While there is no scaling attractor in non-interacting
scenario, by considering an interaction between dark energy and dark
matter, we find scaling attractors in which the fractional densities
of dark energy and matter are non-vanishing constants over there.
The universe can evolve to these scaling attractors regardless of
initial conditions and the cosmological coincidence problem could be
alleviated without fine-tunings.\\


\end{abstract}

\newpage
\section{Introduction}
A mysterious component of our universe is dark energy. It is
responsible for current cosmic acceleration and observationally
confirmed [1-4]. $\Lambda$CDM model is the simplest theoretical
explanation for dark energy in which a cosmological constant
$\Lambda$ accounts for dark energy and it is in best agreement with
the observational data. Besides its successes, $\Lambda$ suffers
from two major problems: the so-called cosmological constant problem
and
the cosmic coincidence problem [5, 6].\\
To solve these problems, one needs to introduce further dynamical
degrees of freedom, either by modifying the right hand side of the
Einstein equations or the left hand side of them. In the first class
usually a scalar field such as quintessence [7], phantom [8],
tachyon [9] and so on, appears in the source sector while in the
second class the Ricci scalar $R$ is replaced by often higher order
terms in the gravitational field. The well-known $f(R)$ [10, 11],
$f(G)$ [12, 13] and $f(R,G)$ [14-16] models belong to this later
class and it is proved that after some transformations and
redefinitions these theories can be recast into scalar-tensor
theories.\\
In scalar-tensor theories such as the Brans-Dick theory [17], scalar
field non-minimally couples to gravity. The non-minimal coupling of
the form $f(\varphi)R$ is motivated by many reasons such as quantum
corrections, renormalizability of the scalar field theory in curved
space, Klauze-Klein compactification scheme, the low-energy limit of
superstring theory etc. The most general non-minimally coupled
scalar field theory results in second order field equations, is
known as the Horndeski model [18] in which higher order derivatives
of
the scalar field appear in its Lagrangian.\\
Furthermore, while dark energy models usually assume the filed to be
non-interacting considering an interaction between different
components of the universe could provide solutions to a number of
cosmological problems such as the coincidence problem via the method
of scaling solution [19-22]. In fact, as nothing specific is known
about the nature of both dark energy and dark matter, there is no
physical reason to exclude the possible interaction between them. In
addition, there are some observational evidences for interaction in
dark sectors of our universe [23, 24]. These facts motivate us to
study a new proposed dark energy model in an interacting scenario.\\
Moreover, an alternative gravitational theory to general relativity
is the teleparallel gravity where gravity is described by using the
Weitzenbock connection (for review see [25]). The Lagrangian of this
theory is written with the torsion scalar $T$, and not the Ricci
scalar $R$ defined with Levi-Civita connection [26-28] as in general
relativity.\\
Recently inspired by the same theories in the framework of general
relativity, various generalization of teleparallel gravity including
$f(T)$ gravity [29-31], $f(T,T_{G})$ gravity [32] as well as
non-minimally coupled scalar-torsion theories [33] have been wildly
studied in the literature (see for example [34, 35]).\\
A non-minimal coupling between derivative of scalar field and
curvature in the framework of general relativity is also another
interesting issue to explore cosmological implications of such a
coupling in details [36-38]. The same non-minimal coupling of the
form $T\,g^{\mu\nu}\,\partial_{\mu}\varphi\partial_{\nu}\varphi$ is
considered in [39] to obtain black hole solution of the model in
teleparallel framework.\\
A new version of the above mentioned scalar-torsion theory has been
proposed in [40, 41] where scalar field through its derivatives
couples to vector part of torsion in the form
$f(\varphi)\,\partial_{\mu}\varphi\,\mathcal{V}^{\mu}$ with
$\mathcal{V}^{\mu}$ as the vector torsion. Recently, Jennen and
Pereira have shown in [54] that such a non-minimal coupling of
scalar field derivative and vector torsion naturally appears in the
context of de Sitter teleparallel gravity which is consistent with
local space time kinematics regulated by the de Sitter group
SO(1,4). In the present work we study the dynamics of such a model
by considering an interaction between dark energy and background
matter. We perform a detailed phase-space analysis of the model for
the most familiar interaction term extensively considered in the
literature. The present
paper is organized as follows:\\
In section 2 we briefly review the teleparallel gravity and then
introduce the model. In section 3 by using suitable auxiliary
variables we build up a system of autonomous differential equations.
In section 4, the critical points of dynamical system are extracted
for constant $\alpha$ (it means the coupling function is linear in
terms of $\varphi$). Fixed points and their stability properties for
a non-constant $\alpha$ have been obtained in section 5. Section 6
is devoted to our conclusion.\\

\section{Teleparallel Gravity and the Model}
Here we briefly review the formulation of teleparallelism. The
metric is written as
$g_{\mu\nu}=\eta_{ab}\,e_{\mu}^{a}\,e_{\nu}^{b}$ where $\eta_{ab}$
is the Minkowski metric, $e_{a}(x^{\mu})$ are orthonormal tetrad
components at points $x^{\mu}$ of the manifold in the tangent space
and $e_{a}^{\mu}$ is the tangent vector of the manifold. As we
mentioned earlier the Lagrangian density is given by torsion scalar
$T$ defined as $T\equiv S_{\rho}\,^{\mu\nu}T^{\rho}\,_{\mu\nu}$,
where $T^{\rho}\,_{\mu\nu}\equiv
e_{a}^{\rho}\big(\partial_{\mu}e^{a}_{\nu}-\partial_{\nu}e^{a}_{\mu}\big)$
is the torsion tensor and $S_{\rho}\,^{\mu\nu}\equiv
\frac{1}{2}\big(K^{\mu\nu}\,_{\rho}+\delta^{\mu}_{\rho}\,T^{\alpha\nu}\,_{\alpha}-\delta^{\nu}_{\rho}\,T^{\alpha\mu}\,_{\alpha}\big)$
with $K^{\mu\nu}\,_{\rho}\equiv
-\frac{1}{2}\big(T^{\mu\nu}\,_{\rho}-T^{\nu\mu}\,_{\rho}-T_{\rho}\,^{\mu\nu}\big)$
is the contorsion tensor.\\
By considering the FRW metric background as
\begin{eqnarray}
ds^{2}=dt^{2}-a^{2}(t)(dr^{2}+r^{2}d\Omega^{2}),
\end{eqnarray}
the metric and tetrad components read
$g_{\mu\nu}=diag(1,-a^{2},-a^{2},-a^{2})$ and
$e^{a}_{\mu}=(1,a,a,a)$ respectively. In FRW background the torsion
scalar is also given by $T=-6H^{2}$. Here, $H=\frac{\dot{a}}{a}$ is
the Hubble parameter, $a$ is the scale factor and a dot stands for
derivative
with respect to the cosmic time $t$.\\
Furthermore, the torsion scalar can be decomposed into three
components namely vector torsion, axial torsion and pure tensor
torsion [25] where by vector torsion we mean
\begin{equation}
\mathcal{V}_{\mu}=T_{\,\,\nu\mu}^{\nu}.
\end{equation}
The Ricci scalar of the Levi-Civita connection in terms of the
Weitzenbock connection can be expressed as
\begin{equation}
R=-T+2\nabla_{\mu}\,\mathcal{V}^{\mu}=-T+\frac{2}{e}\partial_{\mu}\,(e\mathcal{V}^{\mu}),
\end{equation}
where $\nabla_{\mu}$ is the Levi-Civita covariant derivative.\\
Moreover, a non-minimal coupling between a scalar field and the
Ricci scalar of the form $\xi R \varphi^{2}$, because of the
motivations such as quantum corrections to the scalar field in
curved space-time [46-47], renormalization considerations [48] and
its appearance in the context of superstring theories [49] is
completely reasonable.\\
Now, if ones try to construct the same theory in the teleparallel
framework, they should consider a non-minimal coupling between
scalar field and the boundary term (total derivative term in (3)
[50, 51]) of the form $C(\varphi)\nabla_{\mu}\,\mathcal{V}^{\mu}$.
However, integration by parts from such a coupling leads to the
following equivalent coupling
$f(\varphi)\,\mathcal{V}^{\mu}\partial_{\mu}\varphi$ where
$f(\varphi)=C'(\varphi)$ and this is the form of coupling we will
consider in the following.\\
Now let us introduce the model in which scalar field non-minimally
couples to the vector torsion through its derivative
\begin{equation}
 S=\int d^{4}x\,e\,\left[\frac{T}{2\,\kappa^2}-\frac{1}{2}\partial_{\mu}\varphi\partial^{\mu}\varphi-V(\varphi)+\eta\,f(\varphi)
 \partial_{\mu}\varphi\, \mathcal{V}^{\mu}\right]+S_{m},
\end{equation}
where $e\equiv\det(e^{a}_{~\mu})=\sqrt{-g}$, $\kappa^{2}=8\pi G$ and
$G$ is a bare gravitational constant (for reviews on teleparallelism
see [25]). $S_{m}$ is the matter action, $\eta$ is a dimensionless
constant measuring the non-minimal coupling and
$f(\varphi)$ is an general function of the scalar field.\\
The gravitational coupling of fundamental fields in teleparallel
gravity is a controversial subject. Since there is no experimental
data to help us, we should rely on equivalence between general
relativity and teleparallel gravity. According to this formulation,
each one of the fundamental fields of nature (scalar, spinor, and
electromagnetic) are required to couple to torsion in such a way to
preserve the equivalence between teleparallel gravity and general
relativity. It is shown in [25] that in the context of teleparallel
gravity a scalar field by itself does not feel gravity but its
four-derivative (which is a vector field) interacts with the vector
part of the torsion. On the other hand, in the framework of
teleparallelism, due to the presence of a total derivative in the
relation between the curvature scalar and the torsion scalar the
essential condition for the torsion scalar $T$ to has its
non-minimal coupling to a scalar field in a conformal manner is that
a term of the form
$f(\varphi)\partial_{\mu}\varphi\,\mathcal{V}^{\mu}$ has to be
assumed in the action [52]. Also, it has been shown in Ref [53] that
a Lagrangian which is invariant under the space-time coordinate
transformations and conformal transformations and leads to
teleparallel Lagrangian in the gauge where the scalar field is
restricted to assume a constant value, includes a non-minimal
coupling of the form
$g^{\mu\nu}\,\mathcal{V}_{\nu}\varphi\,(\partial_{\mu}\varphi)$.\\
The model (4) and teleparallel dark energy model [33] in which
scalar field non-minimally coupled to the torsion scalar instead of
vector torsion, are mathematically related through a conformal
transformation although they are physically different. In what
follows we consider a possible interaction between dark energy and
dark matter and study the phase-space of the model (4).\\
Varying action (4) with respect to the tetrad field yields to the
field equations. In this procedure the energy momentum tensor
associated with the scalar field $\Theta_{a}\vspace{0.1
mm}^{\rho}\equiv-\frac{1}{e}\frac{\delta S_{\varphi}}{\delta
e^{a}\vspace{0.1 mm}_{\rho}}$ is given by

$$\Theta_{a}\vspace{0.1
mm}^{\rho}=\eta\big[f(\varphi)(\mathcal{V}^{\rho}\partial_{a}\varphi+\nabla_{a}\partial^{\rho}\varphi
-e_{a}\vspace{0.1
mm}^{\rho}\nabla_{\mu}\partial^{\mu}\varphi)+f_{,\varphi}
(\partial_{a}\varphi\,\partial^{\rho}\varphi-e_{a}\vspace{0.1
mm}^{\rho}\partial_{\mu}\varphi\,\partial^{\mu}\varphi)\big]$$
\begin{equation}
-e_{a}\vspace{0.1
mm}^{\rho}\big(\frac{1}{2}\partial_{\mu}\varphi\,\partial^{\mu}\varphi-V(\varphi)\big)+\partial_{a}\varphi\,\partial^{\rho}\varphi,
\end{equation}
where $\nabla^{\mu}$ is the covariant derivative in the teleparallel
connection [25], $f_{,\varphi}= \frac{d f(\varphi)}{d\varphi}$
and $V_{,\varphi}=\frac{dV}{d\varphi}$.\\
By imposing the flat FRW metric in (5), we obtain the Friedmann
equations with scalar field energy density and pressure of the form,
\begin{equation}
 \rho_{\varphi}=\frac{1}{2}\dot{\varphi}^{2}+V(\varphi)-3\eta f(\varphi)H \dot{\varphi},
\end{equation}
and
\begin{equation}
 p_{\varphi}=\frac{1}{2}(1+2\,\eta \,f_{,\varphi})\dot{\varphi}^{2}-V(\varphi)+\eta
 f(\varphi)\ddot{\varphi}.
\end{equation}
Additionally, variation of action (4) with respect to the scalar
field yields to its evolution equation that in FRW background takes
the form
\begin{equation}
 \ddot{\varphi}+3\,H\,\dot{\varphi}-3\,\eta\,\left(\dot{H}+3H^{2}\right)\,f(\varphi)+V_{,\varphi}=-\frac{Q}{\dot{\varphi}},
\end{equation}
where $Q$ is a general interaction term corresponding to coupling
between dark energy and dark matter. In fact equation (8) expresses
the continuity relation for the field
$\dot{\rho}_{\varphi}+3\,H\left(1+\omega_{\varphi}\right)\rho_{\varphi}=-Q$
with $\omega_{\varphi}= p_{\varphi}/\rho_{\varphi}$ the equation of
state of the scalar field while the continuity equation for matter
reads $\dot{\rho}_{m}+3H(1+\omega_{m})\rho_{m}=Q$.\\
Here, we mention that considering an interaction between dark energy
and dark matter is a common way for elaboration of the well-known
coincidence problem or why are the densities of vacuum energy and
dark matter equal today? In dynamical system technique this problem
can be alleviated via the method of scaling solutions [19-22]
(solution corresponding to accelerating universe and ratio
$\frac{\rho_{\varphi}}{\rho_{m}}=c$, with $c$ a non-zero
constant).\\

\section{Dynamical Analysis}
In order to study the phase-space and stability analysis of the
model using the dynamical system method let us introduce the
following auxiliary variables:
\begin{equation}
 x\equiv\frac{\kappa\,\dot{\varphi}}{\sqrt{6}\,H}, \:\:\:\:\:\: y\equiv\frac{\kappa\,\sqrt{V}}{\sqrt{3}\,H}, \:\:\:\:\:\:
   u\equiv\kappa\,f, \:\:\:\:\ \alpha\equiv f_{,\varphi}, \:\:\:\: \lambda\equiv-\frac{V_{,\varphi}}{\kappa V}.
\end{equation}
In terms of these new variables, the field equations can be
rewritten as follows,
\begin{equation}
 \frac{dx}{dN}=(3-s)\Big(-x+\frac{\sqrt{6}}{2}\eta
 u\Big)+\frac{\sqrt{6}}{2}\,\lambda\, y^{2}-\hat{Q},
  \end{equation}
\begin{equation}
 \frac{dy}{dN}=\left(-\frac{\sqrt{6}\,\lambda}{2}\,x\,y+s\right)\,y,
\end{equation}

\begin{equation}
 \frac{du}{dN}=\sqrt{6}\,\alpha\, x,
\end{equation}

\begin{equation}
 \frac{d \lambda}{dN}=-\sqrt{6}\,\lambda^{2}\,x\,\left(\Gamma-1\right),
 \end{equation}

\begin{equation}
 \frac{d \alpha}{dN}=\sqrt{6}\,x\,\Pi,
 \end{equation}
 where $N=\ln{a}$, $\hat{Q}=\frac{\kappa\, Q}{\sqrt{6}\,H^{2}\,\dot{\varphi}}$ and the following
 parameters are defined
\begin{equation}
\Pi=\frac{f_{,\varphi\varphi}}{\kappa}, \:\:\:\:\: \:\:\:\:\:\:
\Gamma=\frac{V\,V_{,\varphi\varphi}}{V_{,\varphi}^2}.
\end{equation}
Also, $s$ in our setup reads,
\begin{multline}
 s=-\frac{\dot{H}}{H^{2}}=3\,\left(\frac{2}{3}+\eta^{2}u^{2}\right)^{-1}\\
\left[2x^{2}(1+\eta\alpha)+2\sqrt{6}\eta x
u+3\eta^{2}u^{2}+\lambda\eta
y^{2}u+\gamma(1-x^{2}-y^{2}+\sqrt{6}\eta u x)-\sqrt{\frac{2}{3}}\eta
u \hat{Q}\right],
\end{multline}
where $\gamma$ is the barotropic index defined by
$\gamma=1+\omega_{m}$ such that $1<\gamma<2$.\\
Using variables (9) the density parameters
$\Omega_{i}\equiv(\kappa^{2}\,\rho_{i})/(3\,H^{2})$ for the scalar
field and background matter are given by
\begin{equation}
 \Omega_{\varphi}=x^{2}+{y}^{2}-\sqrt{6}\,\eta\, u\, x, \:\:\:\:\:\:\:\:
 \Omega_{m}=1-\Omega_{\varphi}.
\end{equation}
Also the equation of state of the field $\omega_{\varphi}$, the
effective equation of state and deceleration parameter $q$ can be
written as
\begin{equation}
 \omega_{\varphi}=\frac{p_{\varphi}}{\rho_{\varphi}}=\frac{(1+2\,\eta\,\alpha)x^{2}-y^{2}+\eta\,u\,\big(-\sqrt{6}x+\eta\,
 u(3-s)+\lambda\,y^{2}-\sqrt{\frac{2}{3}}\hat{Q}\big)}
 {x^{2}+{y}^{2}-\sqrt{6}\,\eta\, u\, x},
 \label{21}
\end{equation}

\begin{multline}
\omega_{eff}=\left(p_{\varphi}+p_{m}\right)/\left(\rho_{\varphi}+\rho_{m}\right)\\
=\left(\gamma-1\right)\big[1-(x^{2}+{y}^{2}-\sqrt{6}\,\eta\, u\,
x)\big]+(1+2\,\eta\,\alpha)x^{2}-y^{2}+\eta\,u\,\big(-\sqrt{6}x+\eta\,
 u(3-s)+\lambda\,y^{2}-\sqrt{\frac{2}{3}}\hat{Q}\big).
\end{multline}
and
\begin{equation}
q\equiv-1-\frac{\dot{H}}{H^{2}}=\frac{1}{2}+\frac{3}{2}\omega_{eff}.
\end{equation}
In what follows we utilize $\omega_{eff}<-\frac{1}{3}$ to obtain
the required conditions for an accelerating universe.\\
At this point we briefly review the dynamical system method and its
autonomous system of equations. An autonomous system in general can
be written as $\frac{d\textbf{Y}}{d \ln a}=f(\textbf{Y})$, where the
column vector $\textbf{Y}$ is constituted by suitable auxiliary
variables and $f(\textbf{Y})$ is the corresponding column vector of
the autonomous equations [42-44].\\
The solutions of the system of differential equations namely the
solutions of $f(\textbf{Y})=0$, yield to the fixed (critical) points
$\textbf{Y}_{c}$ of the system. In order to study the stability of
the equilibrium or critical points we should first expand the system
around $\textbf{Y}_{c}$ as $\textbf{Y}=\textbf{Y}_{c}+\textbf{U}$
where the column vector $\textbf{U}$ denotes the perturbation of the
variables. For each critical point the 1st order perturbation
technique leads to the matrix equation $\textbf{U}'=\Sigma .
\textbf{U}$ where the matrix $\Sigma$ contains all the coefficients
of the perturbation equations. The stability of the critical points
can be obtained from the sign of the eigenvalues of $\Sigma$. If the
real part of all eigenvalues at a fixed point are negative then the
fixed point is a stable point otherwise it is an unstable one or a
saddle (in fact if all the eigenvalues have positive real part, then
the fixed point is an unstable fixed point and if some eigenvalues
have negative real part and remaining eigenvalues have positive real
part, then it is a saddle point). A detailed analysis of the
stability criteria is given in Refs [42-44]. A critical point is an
attractor when it is a stable point and the universe evolves to the
attractor solutions regardless of the
initial conditions.\\
In our setup once the parameters $\Gamma$ and $\Pi$ are known,
equations (10)-(14) become a system of autonomous differential
equations and one can study dynamics of the model in a usual way.
Considering an exponential potential of the form $V=V_{0}
\,e^{-\lambda\kappa\varphi}$ with constant $\lambda$ leads to
$\Gamma=1$ and equation (13) can be eliminated from our system of
differential equations. In the other hand, we classify our study in
two parts. First, we consider a coupling function of the form
$f(\varphi)\propto\varphi$ which yields to a constant $\alpha$.
Consequently for an exponential potential and a constant $\alpha$
our system of autonomous equations is reduced to equations
(10)-(12). In the second step a dynamically changing $\alpha$ is
studied. For this case we mention that $u=\kappa\,f$ is a function
of $\varphi$ and one can express $\varphi$ as a function of $u$
using the inverse function i.e
$\varphi(u)=f^{-1}\big(\frac{u}{\kappa}\big)$. Thus,
$\alpha(\varphi)$ and $\Pi(\varphi)$ can be written in terms of $u$
and the dynamical system described by equations (10)-(12) (for
details see
[45]).\\
In the following sections we extract the critical points and their
properties for constant and dynamically changing $\alpha$. Also the
interaction between dark energy and dark matter is assumed to be of
the form $Q=\beta\,\kappa\,\rho_{m}\,\dot{\varphi}$ which leads to
the following $\hat{Q}$ in equation (10)
\begin{equation}
\hat{Q}=\sqrt{\frac{3}{2}}\,\beta\,\Omega_{m},
\end{equation}
where $\beta$ is a dimensionless constant.\\

\section{Constant $\alpha$}
In this section we assume that $\alpha$ in (9) is a no-zero
constant. This means that the non-minimal coupling function is
proportional to $\varphi$ i.e $f(\varphi)\propto\varphi$. By
inserting (21) in (10), the dynamical system (10)-(12) has two
critical points $A_{1}$ and $A_{2}$ presented in Table 1. This table
also provides the corresponding values of density
$(\Omega_{\varphi})$ and equation of state $(\omega_{\varphi})$
parameters of dark energy as well as the effective equation of state
$(\omega_{eff})$ at each critical points.\\
Now substituting the liner perturbations $x\rightarrow x_{c}+\delta
x$, $y\rightarrow y_{c}+\delta y$ and $u\rightarrow u_{c}+\delta u$
into the autonomous system (10)-(12) and linearize them give us the
components of the perturbation matrix $\Sigma$ (the component of
$\Sigma$ have been written in the Appendix). The sign of the real
part of the eigenvalues of $\Sigma$ will determine the type and
stability of the critical points. The results of stability analysis
have been summarized in Table 2. We have also presented the
existence and acceleration conditions in Table 2. Let us now discuss
the properties of critical points in details.\\

\textbf{Critical Point $A_{1}$:}\\
This point corresponds to a completely dark energy dominated
solution $(\Omega_{\varphi}=1)$ with equation of state equal to the
cosmological constant $(\omega_{\varphi}=1)$. Accelerated expansion
occurs at point $A_{1}$ for all values of the model parameters.
Three eigen values of matrix $\Sigma$ at this point are as follows:
$$\mu_{12}=\frac{3}{2}\Big(-1\pm \sqrt{1+\frac{24\,\eta\,\alpha}{\lambda^{2}+6}}\Big),\,\,\,
\,\,\,\,\,\mu_{3}=-3\gamma.$$ Therefore, $A_{1}$ is a stable point
if $\eta\alpha<0$ and thus it can attract the universe at
late-times.\\
Using numerical computations and phase-space trajectories, we have
shown the attractor behavior of point $A_{1}$
for special choices of the model parameters in Figure 1.\\

\textbf{Critical Point $A_{2}$:}\\
Point $A_{2}$ corresponds to a matter dominated solution that exists
under condition
$2\beta^{2}>12(\gamma-1)-3\gamma^{2}+2(\gamma-2)\gamma\,\eta$. The
deceleration parameter at this point is given by
$q=\frac{3}{2}\gamma-1$. Thus expansion of the universe is
non-accelerating. Eigenvalues of linearized perturbation matrix at
this point are obtained as,
$$\mu_{1,2}=\frac{3}{2}(\gamma-2)\Big(-1\pm \sqrt{1+\frac{2\,\gamma\,\eta\,(2-\gamma)}{(3\gamma^{2}+2\beta^{2}-12\gamma+12)}}\Big),\,\,\,
\,\,\,\,\,\mu_{3}=\frac{3}{2}\gamma.$$ So, $A_{2}$ is an unstable
point because one of the eigenvalues $(\mu_{3})$ is positive. It may
deserve to be considered as a possible state of the universe at
previous stages.\\

\begin{table}[t]
\caption{The critical points of the autonomous system (10)-(12) for
constant $\alpha$ and the corresponding values of the dark energy
density parameter $\Omega_{\varphi}$, the dark energy equation of
state parameter $\omega_{\varphi}$ and the effective equation of
state parameter $\omega_{eff}$.}
 \centering
\begin{center}
\begin{tabular}{|c|c|c|c|c|c|c|}\hline
Name & $x_{c}$& $y_{c}$& $u_{c}$  & $\Omega_{\varphi}$ &
$\omega_{\varphi}$ &$\omega_{eff}$\\
\hline $A_{1}$ & $0$&$1$&$-\frac{\lambda}{3\eta}$ & $1$ & $-1$ &
$-1$\\\hline
 $A_{2}$ &  $0$&$0$&$-\frac{2\beta}{3\eta(\gamma-2)}$ &
$0$ & $-1$ & $\gamma-1$\\\hline
\end{tabular}
\end{center}
\end{table}

 \begin{table}
  \caption{Existence, acceleration and stability conditions of the fixed points in Table 1.}
 \centering
\begin{center}

\begin{tabular}{|c|c|c|c|}

   \hline label & existence & acceleration & stability  \\
  \hline $A_{1}$& $\eta\,\alpha\geq-\frac{\lambda^{2}+6}{24}$&All values&$\eta\,\alpha<0$\\
 \hline $A_{2}$&$2\beta^{2}>12(\gamma-1)-3\gamma^{2}+2(\gamma-2)\gamma\,\eta$&No&Unstable\\
     \hline
\end{tabular}
\end{center}
\end{table}

\section{Varying $\alpha$}
When the non-minimal coupling function is any general function of
$\varphi$ other than $f(\varphi)\propto\varphi$ then the parameter
$\alpha$ will be a non-constant parameter. As we mentioned in
section 3, $\alpha$ can be expressed in terms of $u$ such that at
critical point $(x_{c},y_{c},u_{c})$,
$\alpha(u)\rightarrow\alpha(u_{c})=0$. Also when $(x,y,u)\rightarrow
(x_{c},y_{c},u_{c})$, the field $\varphi$ rolls down toward
$\pm\infty$ with $f(\varphi)\propto\frac{1}{\kappa}$ and $u_{c}=1$.
Five critical points of dynamical system (10)-(12) together with the
corresponding values of $\Omega_{\varphi}$, $\omega_{\varphi}$ and
$\omega_{eff}$ are presented in Table 3. The existence conditions,
acceleration and stability criteria of these points have been shown
in Table 4. Detailed explanation for each point is as follows:\\

\textbf{Critical Point $B_{1}$:}\\
Point $B_{1}$ exists for all values of the model parameters. Dark
energy density parameter $\Omega_{\varphi}$, equation of state
$\omega_{\varphi}$, effective equation of state $\omega_{eff}$ and
deceleration parameter $q$ at this point are given respectively by
\begin{equation}
\Omega_{\varphi}=-\frac{1}{6}\frac{(3\gamma\,\eta+2\beta-6\eta)(-6\beta\,\eta^{2}+3\gamma
\,\eta-2\beta-6\eta)}{(-\beta\,\eta+\gamma-2)^{2}},
\end{equation}

\begin{equation}
\omega_{\varphi}=\frac{-6\beta(\gamma-1)\eta^{2}+(3\gamma^{2}-2\beta^{2}-9\gamma+6)\eta-2\beta}
{(-6\beta\,\eta^{2}+(3\gamma-6)\eta-2\beta)},
\end{equation}

\begin{equation}
\omega_{eff}=\frac{1}{9}\,\frac{-2\beta^{2}+3(-2\gamma+3)\eta\,\beta+3\gamma(\gamma-3)+6}{(-\beta\,\eta+\gamma-2)},
\end{equation}
and
\begin{equation}
q=\frac{1}{2}\,\frac{-2\beta^{2}+2(-3\gamma+4)\eta\,\beta+\gamma(3\gamma-8)+4}{(-\beta\,\eta+\gamma-2)}.
\end{equation}
It is not easy to obtain a simple expression for acceleration
condition $\omega_{eff}<-\frac{1}{3}$ using equation (24). Thus, we
find the acceleration condition only when the matter is
non-relativistic $(\gamma=1)$. In this case under the following
conditions the expansion of the universe is
accelerating \\

\textbf{i)}\,\, if\,$(\beta\,\eta+1)>0$ then we need the parameter
$\beta$ to be satisfied in
\begin{equation}
\frac{\eta}{2}\Big[1-\sqrt{1-\frac{2}{\eta^{2}}}\Big]
<\beta<\frac{\eta}{2}\Big[1+\sqrt{1-\frac{2}{\eta^{2}}}\Big],
\end{equation}
and\\

\textbf{ii)}\,\, if\,$(\beta\,\eta+1)<0$ then we require
$$\beta>\frac{\eta}{2}\Big[1+\sqrt{1-\frac{2}{\eta^{2}}}\Big],$$
or
\begin{equation}
\beta<\frac{\eta}{2}\Big[1-\sqrt{1-\frac{2}{\eta^{2}}}\Big].
\end{equation}
The solution $B_{1}$ can be a scaling solution
$(0<\Omega_{\varphi}<1)$ for particular choices of the model
parameters and hence give hope to alleviate the cosmological
coincidence problem.\\
Three eigenvalues of the corresponding perturbation matrix $\Sigma$
at point $B_{1}$ read

$$\mu_{1}=\frac{-2\beta^{2}-6(\gamma-2)\eta\,\beta+3(\gamma-2)^{2}}{2(-\beta\,\eta+\gamma-2)},$$
$$\mu_{2}=\frac{\tau_{c}(3\gamma\,\eta+2\beta-6\eta)}{(-\beta\,\eta+\gamma-2)},$$
\begin{equation}
\mu_{3}=\frac{-2\beta^{2}-2\big(3(\gamma-1)\eta+\lambda\big)\beta+3(\gamma-2)(-\eta\,\lambda+\gamma)}{2(-\beta\,\eta+\gamma-2)},
\end{equation}
where $\tau_{c}$ in $\mu_{2}$ stands for $\tau_{c}=\frac{d
\alpha(u)}{du}|_{u=u_{c}}$. So, $B_{1}$ could be an attractor
solution for different values of the model parameters.\\
Let us consider the simple case $\gamma=1$ and examine the stability
conditions. In this case $B_{1}$ is a stable point under the
following requirements:\\

\textbf{i)}\,\, if\,$(\beta\,\eta+1)>0$ then one of the required
conditions is as follows
\begin{equation}
-\frac{\lambda}{2}\Big[1+\sqrt{1+\frac{6(\eta\,\lambda-1)}{\lambda^{2}}}\Big]
<\beta<\frac{\lambda}{2}\Big[-1+\sqrt{1+\frac{6(\eta\,\lambda-1)}{\lambda^{2}}}\Big],
\end{equation}
and for $\tau_{c}<0$ the additional condition read
\begin{equation}
\frac{3}{2}\eta\Big[1-\sqrt{1+\frac{2}{3\eta^{2}}}\Big]<\beta<\frac{3}{2}\eta,
\end{equation}
while for $\tau_{c}>0$ in addition to (29) one needs
\begin{equation}
\frac{3}{2}\eta<\beta<\frac{3}{2}\eta\Big[1+\sqrt{1+\frac{2}{3\eta^{2}}}\Big].
\end{equation}

\textbf{ii)}\,\, if\,$(\beta\,\eta+1)<0$ then for $B_{1}$ to be an
stable point one requires
$$\beta>\frac{\lambda}{2}\Big[-1+\sqrt{1+\frac{6(\eta\,\lambda-1)}{\lambda^{2}}}\Big],$$
or
\begin{equation}
\beta<-\frac{\lambda}{2}\Big[1+\sqrt{1+\frac{6(\eta\,\lambda-1)}{\lambda^{2}}}\Big].
\end{equation}
Additionally for $\tau_{c}<0$ we should have
\begin{equation}
\beta>\frac{3}{2}\eta\Big[1+\sqrt{1+\frac{2}{3\eta^{2}}}\Big],
\end{equation}
and for $\tau_{c}>0$
\begin{equation}
\beta<\frac{3}{2}\eta\Big[1-\sqrt{1+\frac{2}{3\eta^{2}}}\Big].
\end{equation}
All in all, point $B_{1}$ could be an attractor scaling solution in
which accelerated expansion of the universe occurs. In the left
panel Figure 2 we have depicted the phase-space trajectories of the
model for special choices of parameters. With these values of
parameters, $B_{1}$ is an attractor point as it is clear from the
figure.\\

\textbf{Critical Point $B_{2}$:}\\
This point is a scalar field dominated solution
$(\Omega_{\varphi}=1)$ but without accelerated expansion because
$\omega_{eff}=1$. Since the eigenvalues of the corresponding matrix
$\Sigma$ at point $B_{2}$ are very complicated, one can not conclude
about its stability analytically. Thus, we should look at the
phase-space trajectories to find whether this point is stable or
not. Our numerical computations show that point $B_{2}$ is a stable
point. Because of the disadvantage that the expansion is not
accelerating, this point is not a realistic solution at late-times
and we leave further explanations about point $B_{2}$.\\

\textbf{Critical Point $B_{3}$:}\\
Point $B_{3}$ has the same properties as point $B_{2}$. It exists
for all values of model parameters, it is a stable dark energy
dominated solution without accelerated expansion. Thus, we don't pay
attention to this point further.\\

\textbf{Critical Point $B_{4}$:}\\
The fixed point $B_{4}$ exists for
$$\beta>\frac{\lambda+3\eta(\gamma-1)}{2}\Big[-1+\sqrt{1-\frac{6(2-\gamma)(\gamma-\eta\,\lambda)}
{\big(\lambda+3\eta(\gamma-1)\big)^{2}}}\Big],$$ or
\begin{equation}
\beta<-\frac{\lambda+3\eta(\gamma-1)}{2}\Big[1+\sqrt{1-\frac{6(2-\gamma)(\gamma-\eta\,\lambda)}
{\big(\lambda+3\eta(\gamma-1)\big)^{2}}}\Big].
\end{equation}
The expressions for $\Omega_{\varphi}$, $\omega_{\varphi}$,
$\omega_{eff}$ and deceleration parameter $q$ at this point are
given by

\begin{equation}
\Omega_{\varphi}=-\frac{1}{2}\frac{3\gamma\,\eta\,\lambda-2\beta^{2}+6\beta\,\eta-2\beta\,\lambda+6\eta\,
\lambda-6\gamma}{(\beta+\lambda)^{2}},
\end{equation}
\begin{equation}
\omega_{\varphi}=(\gamma-1)+\frac{2\gamma\,\beta\,(\beta+\lambda)}{3\eta\big(2\beta+(\gamma+2)\lambda
\big)-2\beta\,(\beta+\lambda)-6\gamma},
\end{equation}
\begin{equation}
\omega_{eff}=\frac{\gamma\,\lambda-\beta-\lambda}{\beta+\lambda},
\end{equation}
and
\begin{equation}
q=\frac{3\gamma\,\lambda-2\beta-2\lambda}{2(\beta+\lambda)}.
\end{equation}
Looking at equation (36), we conclude that point $B_{4}$ is a
scaling solution $(0\leq\Omega_{\varphi}<1)$ for particular values
of parameters $\eta$, $\lambda$, $\beta$ and $\gamma$. Thus this
point can solve the coincidence problem.\\
Let us now examine the stability properties of this point. One of
the eigenvalues of $3\times3$ matrix $\Sigma$ at point $B_{4}$ is
$\mu_{1}=\frac{3\tau_{c}\,\gamma}{\beta+\lambda}$, while two other
eigenvalues are too complicated and hence we do not give their
explicit expressions here. Numerically we find  that if this point
exists, it is stable and thus attract the universe at late-times.
The conditions for stability of point $B_{4}$ is as follows
$$ \tau_{c}>0\,\,\,\,and\,\,\,\, \beta+\lambda<0,$$
or
\begin{equation}
\tau_{c}<0\,\,\,\,and\,\,\,\, \beta+\lambda>0.
\end{equation}
In the right panel of Figure 2 we have chosen the parameters such
that they satisfy the above conditions. One can clearly see the
attractor behavior of point $B_{4}$ in this figure. Another
advantage of this point is that acceleration can be occurred under
the following conditions
$$\beta<\frac{\lambda(3\gamma-2)}{2}\,\,\,\,and\,\,\,\,
\beta+\lambda<0,$$ or
\begin{equation}
\beta>\frac{\lambda(3\gamma-2)}{2}\,\,\,\,and\,\,\,\,
\beta+\lambda>0.
\end{equation}

\textbf{Critical Point $B_{5}$:}\\
Similar to points $B_{2}$ and $B_{3}$, point $B_{5}$ is also a
completely dark energy dominated solution $(\Omega_{\varphi}=1)$
that exists for
$$\eta\,\lambda<2,$$
and
\begin{equation}
-3\big(\eta+\sqrt{\frac{2}{3}+\eta^{2}}\big)<\lambda<3\big(-\eta+\sqrt{\frac{2}{3}+\eta^{2}}\big).
\end{equation}
Dark energy equation of state, effective equation of state and
deceleration parameter at point $B_{5}$ are as follows:
\begin{equation}
\omega_{\varphi}=\frac{1}{3}\,\frac{9\eta\,\lambda+2\lambda^{2}-6}{2-\eta\,\lambda},
\end{equation}
\begin{equation}
\omega_{eff}=\frac{1}{3}\,\frac{9\eta\,\lambda+2\lambda^{2}-6}{2-\eta\,\lambda},
\end{equation}
and
\begin{equation}
q=\frac{4\eta\,\lambda+\lambda^{2}-2}{2-\eta\,\lambda}.
\end{equation}
Using equation (44) the required condition for acceleration
$(\omega_{eff}<-\frac{1}{3})$ can be obtained as
\begin{equation}
-2\eta\Big[1+\sqrt{1+\frac{1}{2\eta^{2}}}\Big]<\lambda
<2\eta\Big[-1+\sqrt{1+\frac{1}{2\eta^{2}}}\Big].
\end{equation}
Three eigenvalues of Jacobian matrix $\Sigma$ at point $B_{5}$ read,
$$\mu_{1}=\frac{6\eta\,\lambda+\lambda^{2}-6}{2-\eta\,\lambda},$$
$$\mu_{2}=\frac{2\tau_{c}(3\eta+\lambda)}{2-\eta\,\lambda},$$
\begin{equation}
\mu_{3}=\frac{2(\lambda^{2}-3\gamma)+3(\gamma+2)\eta\,\lambda+2\beta_{1}(\lambda+3\eta)}{2-\eta\,\lambda}.
\end{equation}
Thus, this point is an attractor (stable) point if
\begin{equation}
-\frac{1}{4}\big(3\eta(\gamma+2)+2\beta\big)\Big[1+\sqrt{1-\frac{48(\beta\,\eta-\gamma)}{\big(3\eta(\gamma+2)+2\beta\big)^{2}}}\Big]<\lambda
<\frac{1}{4}\big(3\eta(\gamma+2)+2\beta\big)\Big[-1+\sqrt{1-\frac{48(\beta\,\eta-\gamma)}{\big(3\eta(\gamma+2)+2\beta\big)^{2}}}\Big].
\end{equation}
Phase-space trajectories of the model for different initial
conditions, have been plotted in Figure 3 such that the attractor
behavior of point $B_{5}$ is transparent. Note however that,
although point $B_{5}$ is an attractor solution of the autonomous
system, it is not a scaling attractor and this is the disadvantage
of point $B_{5}$.\\
The evolution of density parameters $\Omega_{\varphi}$ and
$\Omega_{m}$ have been also depicted in Figure 4. In this figure we
have considered
$f(\varphi)=\frac{1}{\kappa}\big(1+e^{\kappa\varphi}\big)$ such that
$\alpha(u)=-1+u$ and $\tau_{c}=1$. The present epoch $(N=4)$
corresponds to $\Omega_{\varphi}\approx0.68$ and
$\Omega_{m}\approx0.32$.\\

\begin{table}[t]
\caption{The critical points of the autonomous system (10)-(12) for
dynamically changing $\alpha$ and the corresponding values of the
dark energy density parameter $\Omega_{\varphi}$, the dark energy
equation of state parameter $\omega_{\varphi}$ and the effective
equation of state parameter $\omega_{eff}$.}
 \centering
\begin{center}
\begin{tabular}{|c|c|c|c|c|c|c|}\hline
Name & $x_{c}$& $y_{c}$& $u_{c}$  & $\Omega_{\varphi}$ &
$\omega_{\varphi}$ &$\omega_{eff}$\\
\hline $B_{1}$ &
$\frac{\sqrt{6}}{6}\frac{3\eta(\gamma-2)+2\beta}{-\beta\eta+\gamma-2}$&$0$&$1$
& Eq. (22) & Eq. (23) & Eq. (24)\\\hline
 $B_{2}$ &  $\frac{\sqrt{6}}{2}\eta+\sqrt{1+\frac{3\eta^{2}}{2}}$&$0$&$1$ &
$1$ & $1$ & $1$\\\hline $B_{3}$ &
$\frac{\sqrt{6}}{2}\eta-\sqrt{1+\frac{3\eta^{2}}{2}}$&$0$&$1$ & $1$
& $1$ & $1$\\\hline $B_{4}$ &
$\frac{\sqrt{6}}{2}\frac{\gamma}{\beta+\lambda}$&$\frac{\sqrt{\frac{3}{2}(2-\gamma)(\gamma-\eta\lambda)+3\beta\eta(\gamma-1
)+\beta(\beta+\lambda)}}{|\beta+\lambda|}$&$1$ & Eq. (36) & Eq. (37)
& Eq. (38)\\\hline $B_{5}$ &
$\frac{\sqrt{6}}{3}\frac{(3\eta+\lambda)}{(2-\eta\lambda)}$&$\frac{\sqrt{6(1-\eta\lambda)-\lambda^{2}}\sqrt{\frac{2}{3}+\eta^{2}}}{2-\eta\lambda}$&$1$
& $1$ &Eq. (43) &Eq. (44)
\\\hline
\end{tabular}
\end{center}
\end{table}

 \begin{table}
  \caption{Existence, acceleration and stability conditions of the fixed points in Table 3.}
 \centering
\begin{center}

\begin{tabular}{|c|c|c|c|}

   \hline label & existence & acceleration & stability  \\
  \hline $B_{1}$& All values&Eqs. (26,27)&Eqs. (29)-(34)\\
 \hline $B_{2}$&All values&No&$\begin{array}{c}
                      see\,\, explanations\,\, about\\
this\,\, point\,\, in\,\, the\,\, text\\
                      \end{array}$ \\
     \hline $B_{3}$&All values&No& $\begin{array}{c}
                      see\,\, explanations\,\, about\\
this\,\, point\,\, in\,\, the\,\, text\\
                      \end{array}$ \\
         \hline $B_{4}$ &$\begin{array}{c}
                      \beta>\frac{\lambda+3\eta(\gamma-1)}{2}\Big[-1+\sqrt{1-\frac{6(2-\gamma)(\gamma-\eta\,\lambda)}
                      {\big(\lambda+3\eta(\gamma-1)\big)^{2}}}\Big] \\
                        or\\
                       \beta< -\frac{\lambda+3\eta(\gamma-1)}{2}\Big[1+\sqrt{1-\frac{6(2-\gamma)(\gamma-\eta\,\lambda)}
                      {\big(\lambda+3\eta(\gamma-1)\big)^{2}}}\Big]\\
                      \end{array}$&Eq. (41)&Eq. (40)\\

         \hline $B_{5}$ &$\begin{array}{c}
         \eta\,\lambda<2\\
         and\\
         -3\big(\eta+\sqrt{\frac{2}{3}+\eta^{2}}\big)<\lambda<3\big(-\eta+\sqrt{\frac{2}{3}+\eta^{2}}\big)\\
\end{array}$
         &Eq. (46)&Eq. (48)\\

 \hline

\end{tabular}
\end{center}
\end{table}

\section{Conclusion}
Recently Otalora [40] has proposed a new teleparallel dark energy
model in which the scalar field, responsible for dark energy,
couples to vector torsion through its derivative. The model (4) is
similar to the de Sitter teleparallel gravity, a theory consistent
with local space-time kinematics regulated by the de Sitter group,
proposed in Ref [54]. In such a theory gravitational sector modeled
by teleparallel gravity interacts with the cosmological function due
to a non-minimal coupling between the trace of the covariant
derivative of the vierbein and a non-constant cosmological function
$\Lambda$. Because the cosmological function is not restricted to be
constant, its value can evolve during the cosmological evolution and
thus may be suitable to describe the evolution of the universe from
inflation to dark energy. As it is mentioned in [54] a huge
cosmological term can drive inflation at the early universe and
afterwards, it should decay to a small value to allow the structure
formation of the universe. Then its value should somehow increase to
account for the
late-time accelerated expansion of the universe [60].\\
Additionally a conformally invariant extension of teleparallel
gravity in which derivative of a scalar field non-minimally coupled
to the vector torsion can realize a power-law or the de Sitter
expansion of the universe and also can give rise to the $\Lambda$CDM
model as it was shown in [52]. Later we studied the same model using
a non-canonical scalar field (tachyon) instead of quintessence in
the action [41]. Here we have generalized the Otalora$^{,}$s model
by considering an interaction between dark sectors. Our basic goal
was to examine whether there exist late-time scaling attractors,
corresponding to a accelerated universe or not. Scaling attractors
are those solutions that both density parameters of dark energy
$(\Omega_{DE})$ and dark matter $(\Omega_{DM})$ are non-vanishing
constants over there. Such a solutions can give hope to alleviate
the cosmological coincidence problem and if they posses acceleration
expansion, basic observational requirements will be satisfied.\\
Using dynamical system method we have extracted the critical points
of the model for both constant and non-constant $\alpha$. When
$\alpha$ is constant (it means the coupling function
$f(\varphi)\propto\varphi$) we find two critical point $A_{1}$ and
$A_{2}$ in Table 1. Neither $A_{1}$ nor $A_{2}$ are scaling
attractors. For non-constant $\alpha$ there are five critical points
presented in Table 3. In this case point $B_{1}$ and $B_{4}$ are
scaling attractors under certain conditions for the free parameters
of the model and they have the possibility to explain an accelerated
universe. These two points make the main difference between
interacting and non-interacting scenario at hand. While there is no
scaling attractor in non-interacting scenario, one can switch on the
interaction and obtain such solutions. The universe evolves
to these attractor solutions for different initial conditions.\\
Before closing this section we mention some additional points: the
post-Newtonian limit of teleparallel gravity with a non-minimally
coupled scalar field to the scalar torsion has been investigated in
[55, 56] and it is shown that the model is compatible with scalar
system tests. Spherically symmetric solutions of such a model were
also studied in the literature (see for example [57-59]). In the
case of our model for dark energy the action (4) should passes
standard solar tests i.e Newton law non-violation and possibility to
make spherical solution. So, it would be worthy to study the solar
system constraint on the model parameter and also examine the black
hole and spherical body solutions to obtain additional information
on its novel features. Such projects, although necessary, lie
outside the aim of the present paper and are left for future
investigations.

\begin{figure}[htp]
\begin{center}
\includegraphics{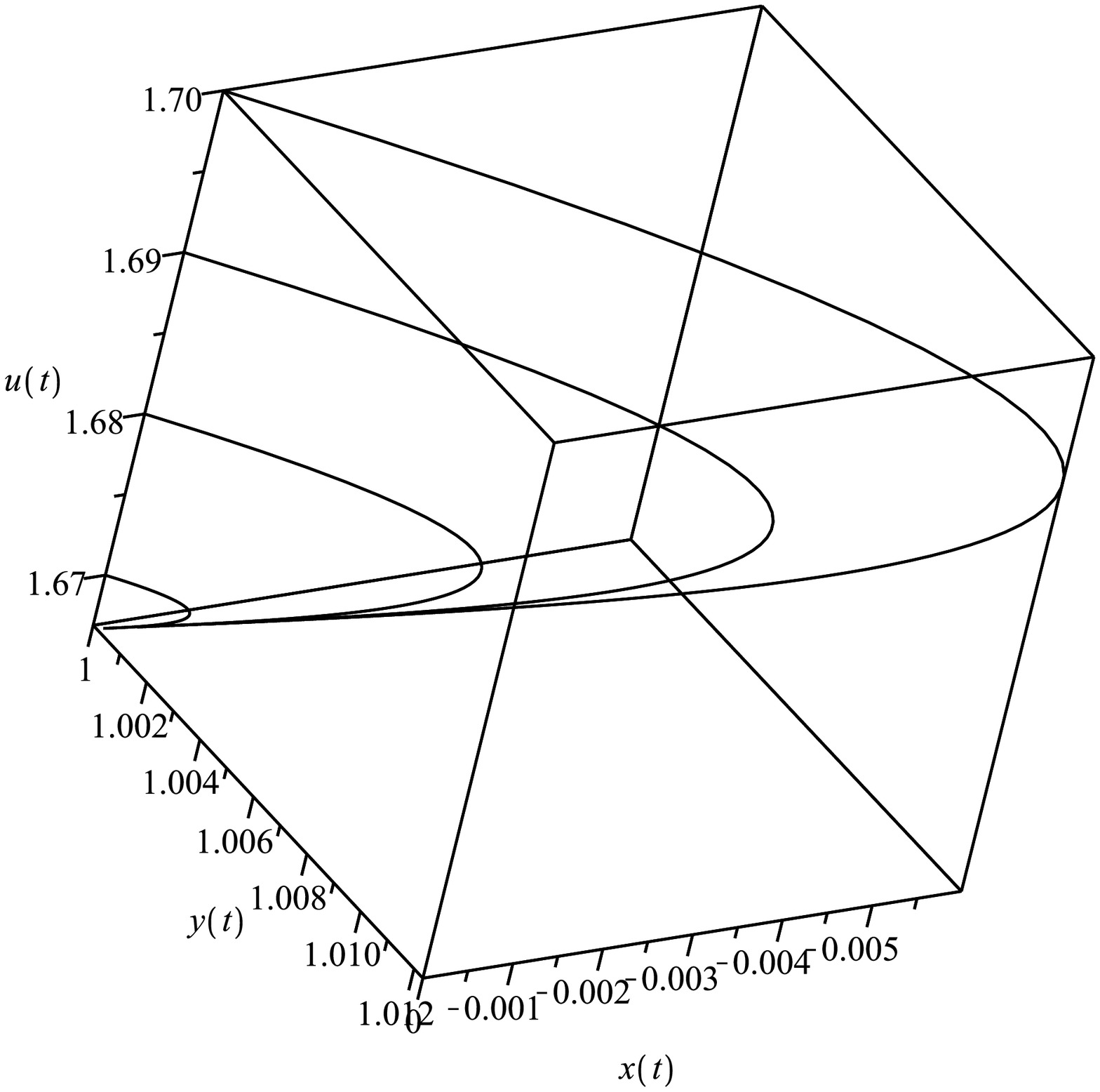} \vspace{2.5cm}\includegraphics{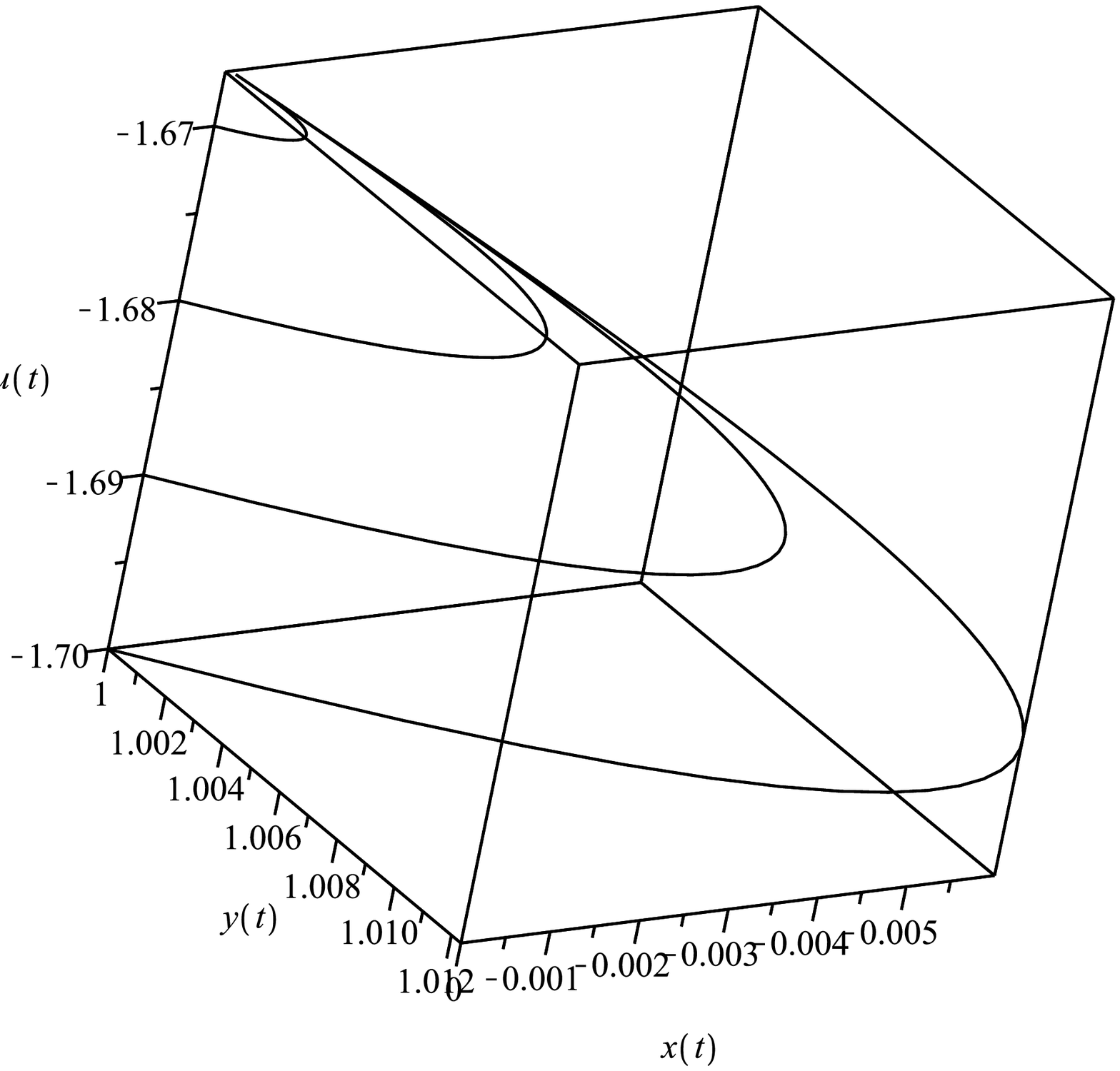}\vspace{4cm}

\caption{\small {3-dimensional phase-space trajectories of the
cosmological scenario (10)-(12) with stable attractor $A_{1}$ for
the parameter choices $\lambda=5$, $\alpha=1$, $\eta=-1$ (left) and
for $\lambda=5$, $\alpha=-1$, $\eta=1$ (right).}}
\end{center}
\end{figure}
\vspace{3cm}
\begin{figure}[htp]
\begin{center}
\includegraphics{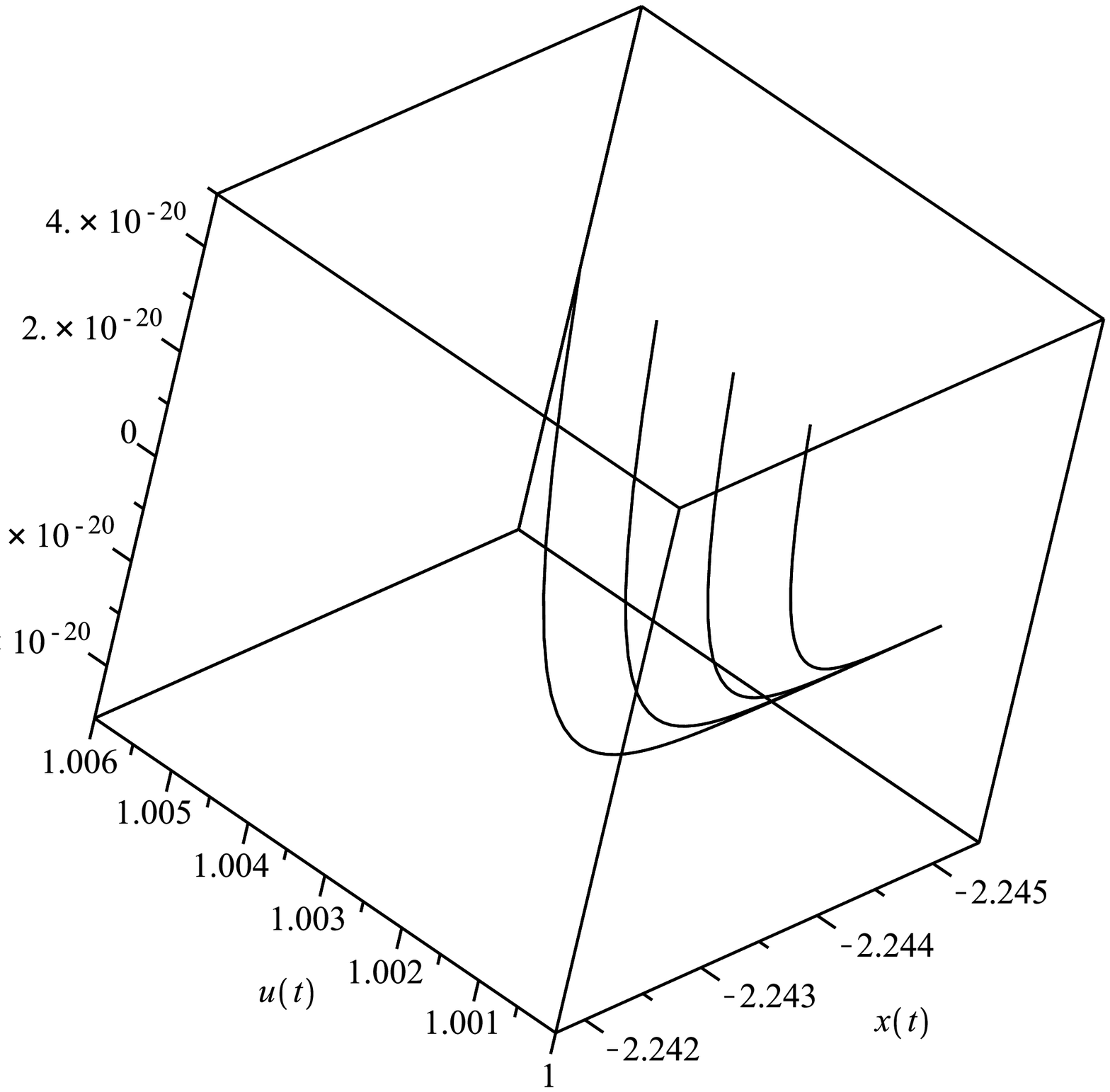} \vspace{2.5cm}\includegraphics{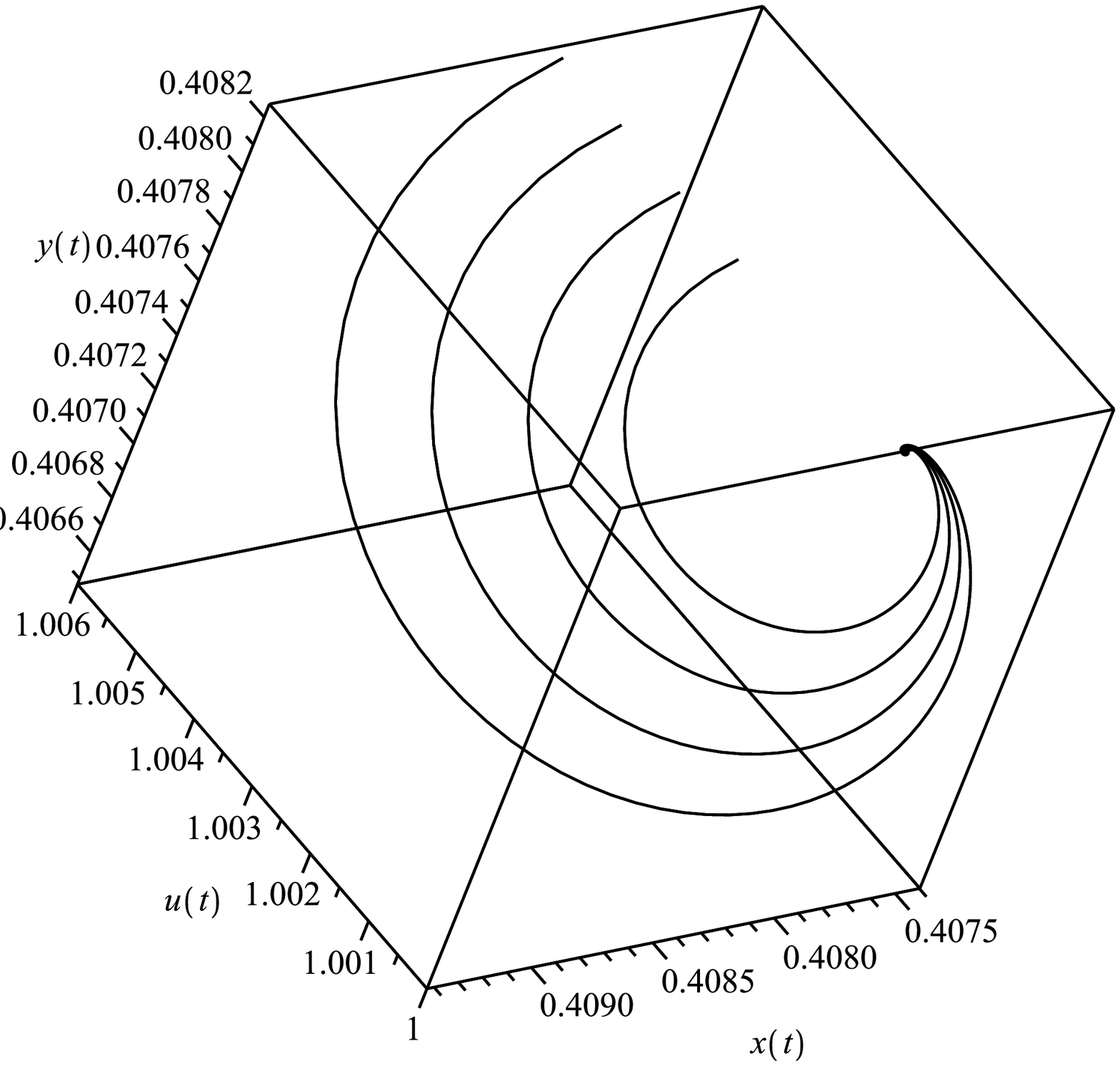}\vspace{2.5cm}

\caption{\small {3-dimensional phase-space trajectories of the
cosmological scenario (10)-(12) with stable attractor $B_{1}$ (left)
for the parameter choices $\gamma=1$, $\lambda=-3$, $\alpha=-1+u$,
$\eta=-1$, $\tau_{c}=1$, $\beta=\frac{1}{3}$ and stable attractor
$B_{4}$ (right) for $\gamma=1$, $\lambda=2$, $\alpha=1-u$, $\eta=1$,
$\tau_{c}=-1$, $\beta=1$. }}
\end{center}
\end{figure}
\vspace{3cm}

\begin{figure}[htp]
\begin{center}
\includegraphics{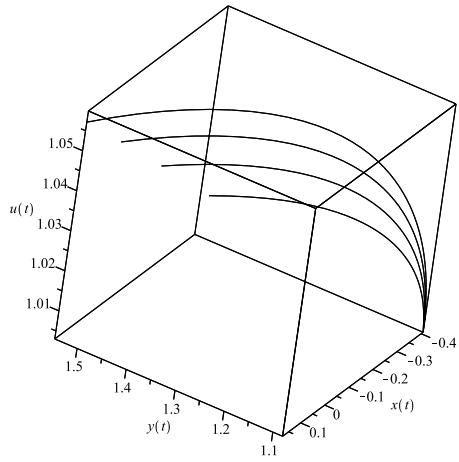} \vspace{7cm}

\caption{\small {3-dimensional phase-space trajectories of the
cosmological scenario (10)-(12) with stable attractor $B_{5}$ for
the parameter choices $\gamma=1$, $\lambda=-2$, $\alpha=1-u$,
$\eta=2$, $\tau_{c}=-1$, $\beta=-1$.}}
\end{center}
\end{figure}
\vspace{3cm}
\begin{figure}[htp]
\begin{center}
\includegraphics{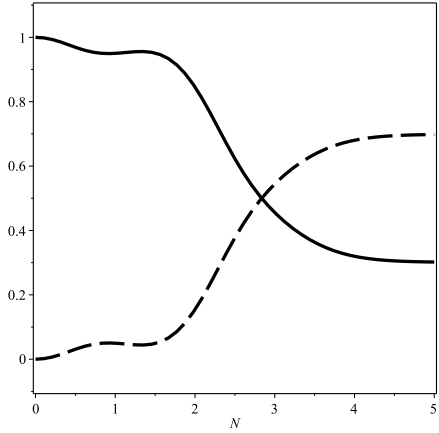} \vspace{5cm}

\caption{\small {Evolution of $\Omega_{\varphi}$ (dotdashed) and
$\Omega_{m}$ (solid) with $\gamma=1$, $\lambda=-3$, $\eta=-1$,
$\tau_{c}=1$, $\beta=\frac{1}{3}$. The initial conditions are
$x_{i}=10^{-8}$, $y_{i}=3.7 \times 10^{-2}$ and $u_{i}=10^{-7}$. The
corresponding values of $\Omega_{\varphi}$ and $\Omega_{m}$ at the
present epoch ($N\simeq 4$) are $\Omega_{\varphi}\approx0.68$ and
$\Omega_{m}\approx0.32$.}}
\end{center}
\end{figure}

\end{document}